\documentclass{article}
\usepackage{eurosym}
\usepackage{amssymb}
\usepackage{amsmath}
\usepackage{cite}
\usepackage{lscape}
\usepackage{epstopdf}
\usepackage{graphicx}
\usepackage{epsf}

\input{tcilatex}

\begin{document}

\title{Periodic solutions from Lie symmetries for the generalized
Chen-Lee-Liu equation}
\author{Andronikos Paliathanasis\thanks{%
Email: anpaliat@phys.uoa.gr} \\
{\ \textit{Institute of Systems Science, Durban University of Technology }}\\
{\ \textit{PO Box 1334, Durban 4000, Republic of South Africa}}\\
{\ \textit{Instituto de Ciencias Fisicas y Matematicas,}}\\
{\ \textit{Universidad Austral de Chile, Valdivia 5090000, Chile}}\\
}
\maketitle

\begin{abstract}
The nonlinear generalized Chen-Lee-Liu 1+1 evolution equation which
describes the propagation of an optical pulse inside a monomode fiber is
studied by using the method of Lie symmetries and the singularity analysis.
Specifically, we determine the Lie point symmetries of the Chen-Lee-Liu
equation and we reduce the equation by using the Lie invariants in order to
determine similarity solutions. The solutions that we found have periodic
behaviour and describe optical solitons. Furthermore, the singularity
analysis is applied in order to write algebraic solutions of the
Chen-Lee-Liu with the use of Laurent expansions. The latter analysis support
the result for the existence of periodic behaviour of the solutions. \newline
\newline
\newline

\end{abstract}

\section{Introduction}

\label{sec1}

Lie symmetry analysis is a powerful method for the study of nonlinear
differential equations. The main feature of Sophus Lie approach is that the
existence of a symmetry vector for a given differential equation indicates
the existence of an invariant surface which can be applied for the
construction of a similarity transformation in order to simplify the
differential equation under the so-called reduction process \cite%
{Bluman,Stephani,olver,ibra}. In addition, Lie symmetries can be used to
determine algebraic equivalent systems \cite{Ovsi} and give linearization
criteria for nonlinear differential equations \cite{lin1,lin2,lin3}.
Furthermore, Lie symmetries can be applied to construct conservation laws 
\cite{con1,con2,con3}; to determine new solutions from old solutions \cite%
{ne1} and many other \cite{ibra}. There are various important results in the
literature on the application of Lie's theory in mathematical physics and
applied mathematics \cite%
{ls01,ls02,ls03,ls04,ls05,ls06,ls07,ls08,ls09,ls10,ls11,ls12,ls13,ls14}.

On the other hand, singularity analysis is an alternative way to study the
integrability of nonlinear differential equations The main requirement for
the singularity analysis is the existence of movable singularities in the
differential equation. Singularity analysis is associated with the French
school led by Painlev\'{e} \cite{pan2,pan3,pan4} and their approach was
actually inspired by the successful application to the determination of the
third integrable case of Euler's equations for a spinning top by
Kowalevskaya \cite{kowa}. For modern approaches of the singularity analysis
we refer the reader in \cite{sin1,sin2,sin3,sin4} and references therein,
while some applications on partial differential equations can be found in 
\cite{ptpde1,ptpde2,ptpde3}.

In this work we are interesting on the determination of exact and analytic
solutions for the generalized Chen-Lee-Liu equation (gCLL) is \cite{gcll1} 
\begin{equation}
iq_{t}+\frac{1}{2}q_{xx}-\left\vert q\right\vert ^{2}q+i\delta \left\vert
q\right\vert ^{2}q_{x}=0,  \label{gc.01}
\end{equation}%
with the application of the Lie's theory and of the singularity analysis.
Equation (\ref{gc.01}) under the Madelung transformation~$q\left( t,x\right)
=\rho \left( t,x\right) \exp \left( i\int u\left( t,x^{\prime }\right)
dx^{\prime }\right) $ the latter equation can be written as the following
system of evolution equations%
\begin{eqnarray}
\rho _{t}+\left( \rho u+\frac{1}{2}\delta \rho ^{2}\right) _{x} &=&0,
\label{gc.02} \\
u_{t}+uu_{x}+\rho _{x}+\delta \left( \rho u\right) _{x}+\left( \frac{\rho
_{x}^{2}}{8\rho ^{2}}-\frac{\rho _{xx}}{4\rho }\right) _{x} &=&0,
\label{gc.03}
\end{eqnarray}%
in which $\rho \left( t,x\right) $ is the intensity variable and chirp
variables, while the positive parameter $\delta \ $is associated with the
self-steepening phenomena \cite{rog1}. The nonlinear evolution equation, has
been proposed for the description of an optical pulse inside a monomode
fiber. For other applications of the equation (\ref{gc.01}) we refer the
reader in \cite{app1}. The solitons provided by the gCLL equation has been
found to be essential for the description of phenomena in optical fiber
theory \cite{gcll1,oss1,oss2,oss3}, while some experimental evidences are
presented in \cite{oss4},

In the following Sections we determine the Lie point symmetries for the real
function system (\ref{gc.02}), (\ref{gc.03}). For the admitted Lie point
symmetries we constructed the algebraic structure and we find the
one-dimensional optimal system. Moreover, we find all possible similarity
transformations which are used to reduce the differential equation and write
the equivalent system. We find four independent similarity solutions. In
particular, we determine static solutions, stationary solutions, travel-wave
solutions and scaling solutions. We observe that the solution are periodic
in the similarity variables which indicate the existence of kink solutions
in the original variables. Furthermore, we apply the singularity analysis
such that to write the analytic solution of the resulting system with the
use of Laurent expansions. While the original system (\ref{gc.02}), (\ref%
{gc.03}) is investigated if possess the Painlev\'{e} property. At this point
it is important to mention some previous studies of Lie's theory on optical
physics \cite{gl01,gl02,gl03}, while\ only recently the algebraic properties
of the Chen-Lee-Liu were studied in \cite{gl10}. The plan of the paper is as
follows.

In Section \ref{sec3}, we determine the Lie symmetries and the similarity
transformations for the gCLL equations, while we construct the similarity
solutions. In Section \ref{sec4} we show that the gCLL equation possess the
Painlev\'{e} property and we write the generic solution by using a Right
Painlev\'{e} Series. Finally, in Section \ref{sec5} we discuss our results
and we draw our conclusions. In Appendices \ref{ap1}, \ref{ap2} and \ref{ap3}
we present the main mathematical theory of the tools that are applied in
this work.

\section{Lie symmetries and similarity transformations}

\label{sec3}

For the system of the 1+1 evolution partial differential equations (\ref%
{gc.02}), (\ref{gc.03}) we apply the Lie theory in order to determine the
generator \cite{ibra,Bluman,Stephani} 
\begin{equation}
X=\xi ^{t}\left( t,x,\rho ,u\right) \partial _{t}+\xi ^{x}\left( t,x,\rho
,u\right) \partial _{x}+\eta ^{\rho }\left( t,x,\rho ,u\right) \partial
_{\rho }+\eta ^{u}\left( t,x,\rho ,u\right) \partial _{u},
\end{equation}%
of the infinitesimal one parameter point transformation%
\begin{equation}
t^{\prime }=t+\varepsilon \xi ^{t}\left( t,x,\rho ,u\right) ~,~x^{\prime
}=x+\varepsilon \xi ^{x}\left( t,x,\rho ,u\right) ~,
\end{equation}%
\begin{equation}
\rho ^{\prime }=\rho +\varepsilon \eta ^{\rho }\left( t,x,\rho ,u\right)
~,~u^{\prime }=u+\varepsilon \eta ^{u}\left( t,x,\rho ,u\right) ~,
\end{equation}%
which keeps invariant the system (\ref{gc.02}), (\ref{gc.03}).

The possible generators are derived to be the following three%
\begin{equation}
X_{1}=\partial _{t}~,~X_{2}=\partial _{x}~\text{and }X_{3}=2t\partial
_{t}+\left( \delta x-t\right) \partial _{x}-\rho \partial _{\rho }-\left(
1+u\delta \right) \partial _{u}.
\end{equation}

The latter Lie point symmetries form the $A_{3,2}$ Lie algebra in the
Morozov-Mubarakzyanov Classification Scheme \cite%
{Morozov58a,Mubarakzyanov63a,Mubarakzyanov63b,Mubarakzyanov63c}. The
commutators and the adjoint representation of the admitted Lie point
symmetries are presented in Tables \ref{com1} and \ref{com2}.

In order to proceed with the derivation of all the possible independent
similarity transformations which reduce the system (\ref{gc.02}), (\ref%
{gc.03}) we should determine the one-dimensional optimal system.
Straightforward, from Table \ref{com2} we find the one-dimensional optimal
system \cite{olver}%
\begin{equation}
\left\{ X_{1}\right\} ~,~\left\{ X_{2}\right\} ~,~\left\{ X_{1}+\alpha
X_{2}\right\} \text{ and }\left\{ X_{3}\right\} \text{.}
\end{equation}

Thus we continue our analysis by applying the Lie point symmetries in order
to reduce the system of partial differential equations (\ref{gc.02}), (\ref%
{gc.03}) into a system of ordinary differential equations.

\begin{table}[tbp] \centering%
\caption{Commutators of the admitted Lie symmetries}%
\begin{tabular}{c|ccc}
\hline\hline
$\left[ \mathbf{X}_{I},\mathbf{X}_{J}\right] $ & $\mathbf{X}_{1}$ & $\mathbf{%
X}_{2}$ & $\mathbf{X}_{3}$ \\ \hline
$\mathbf{X}_{1}$ & $0$ & $0$ & $2X_{1}-X_{2}$ \\ 
$\mathbf{X}_{2}$ & $0$ & $0$ & $X_{2}$ \\ 
$\mathbf{X}_{3}$ & $\,-2X_{1}+X_{2}$ & $-2X_{2}$ & $0$ \\ \hline\hline
\end{tabular}%
\label{com1}%
\end{table}%

\begin{table}[tbp] \centering%
\caption{Adjoint representation of the admitted Lie algebra}%
\begin{tabular}{c|ccc}
\hline\hline
$Ad\left( e^{\left( \varepsilon \mathbf{X}_{i}\right) }\right) \mathbf{X}%
_{j} $ & $\mathbf{X}_{1}$ & $\mathbf{X}_{2}$ & $\mathbf{X}_{3}$ \\ \hline
$\mathbf{X}_{1}$ & $X_{1}$ & $X_{2}$ & $2\varepsilon \delta
X_{1}-\varepsilon X_{2}+X_{3}$ \\ 
$\mathbf{X}_{2}$ & $X_{1}$ & $X_{2}$ & $\varepsilon \delta X_{2}+X_{3}$ \\ 
$\mathbf{X}_{3}$ & $e^{-2\delta \varepsilon }X_{1}-\frac{e^{-\delta
\varepsilon }\left( e^{-\delta \varepsilon }-1\right) }{\delta }X_{2}$ & $%
e^{-\delta \varepsilon }X_{2}$ & $X_{3}$ \\ \hline\hline
\end{tabular}%
\label{com2}%
\end{table}%

\subsection{Similarity transformations}

We proceed with the application of the similarity transformations.

\subsubsection{Lie symmetry $X_{1}$}

The application of the Lie point symmetry $X_{1}$ leads to the static
solution $u=u\left( t\right) ,~\rho =\rho \left( t\right) $ where the
reduced equations are determined to be 
\begin{equation}
\left( \rho u+\frac{1}{2}\delta \rho ^{2}\right) _{,x}=0~,  \label{gc.04}
\end{equation}%
\begin{equation}
uu_{x}+\rho _{x}+\delta \left( \rho u\right) _{x}+\left( \frac{\rho _{x}^{2}%
}{8\rho ^{2}}-\frac{\rho _{xx}}{4\rho }\right) _{x}=0~,  \label{gc.05}
\end{equation}%
or equivalently%
\begin{equation}
\rho u+\frac{1}{2}\delta \rho ^{2}=c_{1},  \label{gc.06}
\end{equation}%
\begin{equation}
\frac{1}{2}u^{2}+\rho +\delta \left( \rho u\right) +\left( \frac{\rho
_{x}^{2}}{8\rho ^{2}}-\frac{\rho _{xx}}{4\rho }\right) =c_{2}~,
\label{gc.07}
\end{equation}%
where now $c_{1},~c_{2}$ are two integration constants. \ 

Hence, from (\ref{gc.05}) it follows 
\begin{equation}
u=-\frac{1}{2}\delta \rho ^{2}+\frac{c_{1}}{\rho },  \label{gc.08}
\end{equation}%
that is, equation (\ref{gc.07}) reads 
\begin{equation}
\varrho _{xx}+\frac{3}{4}\delta ^{2}\varrho ^{5}-2\varrho ^{3}+C_{2}\varrho -%
\frac{c_{1}^{2}}{\varrho ^{3}}=0~,  \label{gc.09}
\end{equation}%
where $\rho =\varrho ^{2}\left( x\right) $ and $C_{2}=\left( \frac{c_{2}}{2}%
-\delta c_{1}\right) $.

Equation (\ref{gc.09}) admits one Lie point symmetry, the vector field $%
X_{2} $, which is a reduced symmetry vector. Moreover, equation (\ref{gc.09}%
) can be integrated by quadratures, that is 
\begin{equation}
\frac{1}{2}\left( \varrho _{x}\right) ^{2}-\frac{\delta }{8}\varrho ^{6}-%
\frac{1}{4}\varrho ^{4}+\frac{1}{2}\left( \frac{c_{2}}{2}-\delta
c_{1}\right) \varrho ^{2}+\frac{c_{1}^{2}}{2\varrho ^{2}}=c_{3}~,
\label{gc.10}
\end{equation}%
where $c_{3}$ is a third integration constant.

\begin{figure}[tbp]
\centering\includegraphics[width=0.8\textwidth]{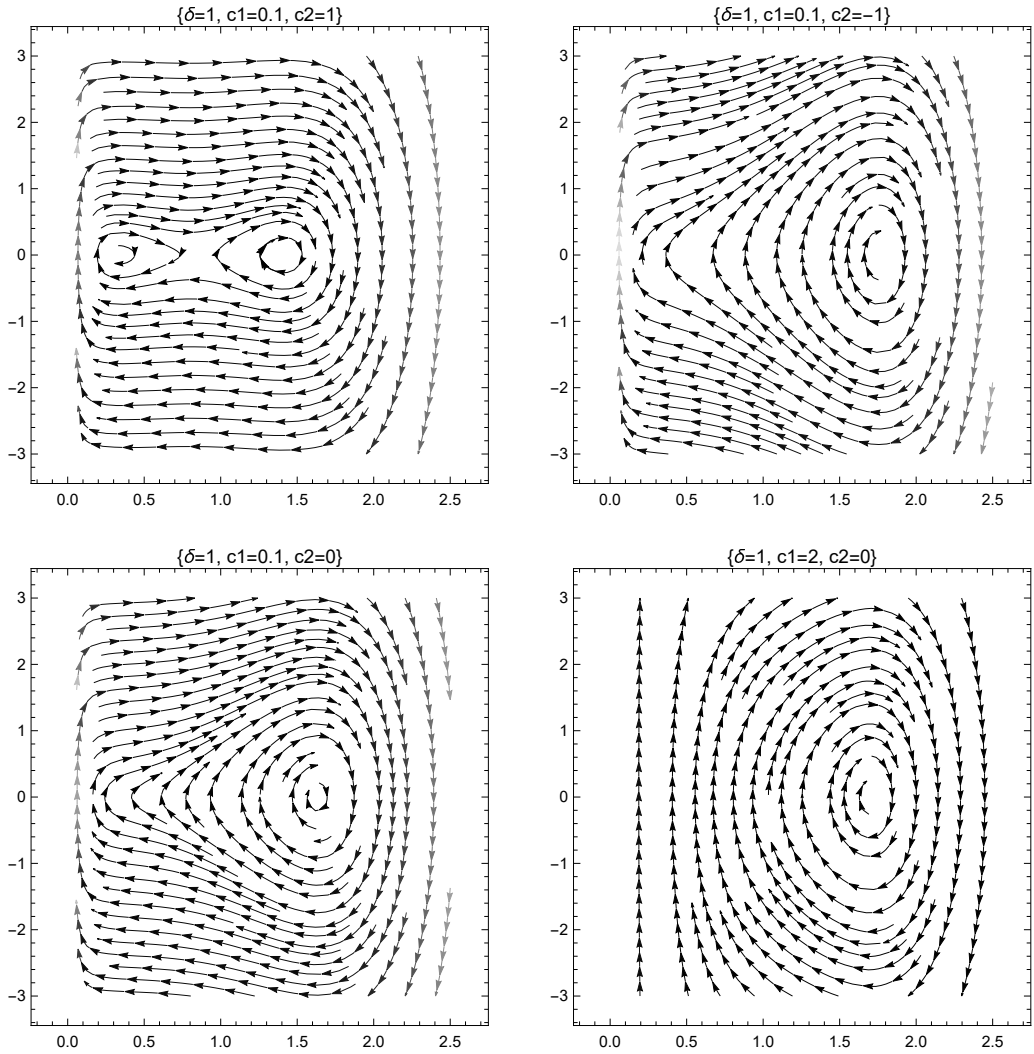}
\caption{Phase portrait for equation (\protect\ref{gc.09}) for various
values of the free parameters $\left\{ \protect\delta ,c_{1},C_{2}\right\} $%
. }
\label{fig1}
\end{figure}

In Fig. \ref{fig1} we present the phase portrait of equation (\ref{gc.09})
from where it is clear that there are attractors in the dynamical system
which provide a periodic behaviour.

In order to write the analytic solution of equation (\ref{gc.09}) and
understand the periodic behaviour of the solution, we apply the singularity
analysis. Specifically we apply the ARS algorithm. We search for the
singular behaviour $\varrho \left( x\right) =\varrho _{0}x^{p}$ from where
we follows that the leading-order behaviour has $p=-\frac{1}{2}$ and $\left(
\varrho _{0}\right) ^{2}=\pm \frac{i}{\delta },$ it is clear from here that
a periodic behaviour will follow.

The second step of the ARS algorithms is the determination of the
resonances, we do that by replacing $\varrho \left( x\right) =\varrho
_{0}x^{-\frac{1}{2}}+mx^{-\frac{1}{2}+S}$ in (\ref{gc.09}) and linearize
around $m=0$, it follows the algebraic equation $\left( S+1\right) \left(
S-3\right) =0$, which gives the resonances $S=-1$ and $S=3$.

Finally we write the Laurent expansion%
\begin{equation}
\varrho \left( x\right) =\varrho _{0}x^{-\frac{1}{2}}+\varrho _{1}x^{\frac{1%
}{2}}+\varrho _{2}x^{\frac{3}{2}}+...~,  \label{gc.11}
\end{equation}%
from where we test that it is a solution of equation (\ref{gc.07}) with
integration constant the $\varrho _{3}$ and%
\begin{equation*}
\left( \varrho _{0}\right) ^{2}=\pm \frac{i}{\delta }~,~\varrho _{1}=-\frac{%
\left( -1\right) ^{\frac{3}{4}}}{2\delta ^{\frac{3}{2}}}~,~\varrho _{2}=-%
\frac{\left( -1\right) ^{1/4}\left( 9-8\bar{C}_{2}\delta \right) }{24\delta
^{\frac{5}{2}}},~...~.
\end{equation*}%
We remark that the second integration constant is the position of the
singularity $x_{0}$.

\subsubsection{Lie symmetry $X_{2}$}

Reduction with respect to the vector field $X_{2}$ provides the stationary
solutions $u=u\left( t\right) ,~\rho =\rho \left( t\right) $ where $u_{t}=0$
and $\rho _{t}=0$, that is $u\left( t,x\right) =u_{0},~\rho \left(
t,x\right) =u_{0}$.

\subsubsection{Lie symmetry $X_{1}+\protect\alpha X_{2}$}

The application of the Lie symmetry vector $X_{1}+\alpha X_{2}$ provides a
travel-wave solution. Indeed we find $u=u\left( z\right) ,~\rho =\varrho
^{2}\left( z\right) $ with $z=x-\alpha t$, where now the reduced system is 
\begin{equation}
\left( u-\alpha \right) \varrho ^{2}+\frac{1}{2}\delta \varrho ^{4}=c_{1}~,
\label{gc.12}
\end{equation}%
\begin{equation}
\varrho _{zz}+2\left( \alpha -\delta \varrho ^{2}\right) \rho u-u^{2}\varrho
-\varrho ^{3}+c_{2}\varrho =0~,  \label{gc.13}
\end{equation}%
where $c_{1},~c_{2}$ are two integration constants.

With the use of (\ref{gc.12}), equation (\ref{gc.13}) reads as%
\begin{equation}
\varrho _{zz}+\frac{3}{4}\delta ^{2}\varrho ^{5}-2\left( 1+\alpha \delta
\right) \varrho ^{3}+\bar{C}_{2}\varrho -\frac{c_{1}^{2}}{\varrho ^{3}}=0~,
\label{gc.14}
\end{equation}%
with $\bar{C}_{2}=\left( c_{2}-\alpha -\delta c_{1}\right) $.

Equation (\ref{gc.14}) is of the form of the static solution (\ref{gc.09})
and can be integrated by quadratures, the only main difference is the
coefficient of the $\varrho ^{3}$, where now it is $\left( 1+\alpha \delta
\right) $. The latter quantity can be positive or negative, either if the
travel wave travel of the left to the right or from the right to the left.

\begin{figure}[tbp]
\centering\includegraphics[width=0.8\textwidth]{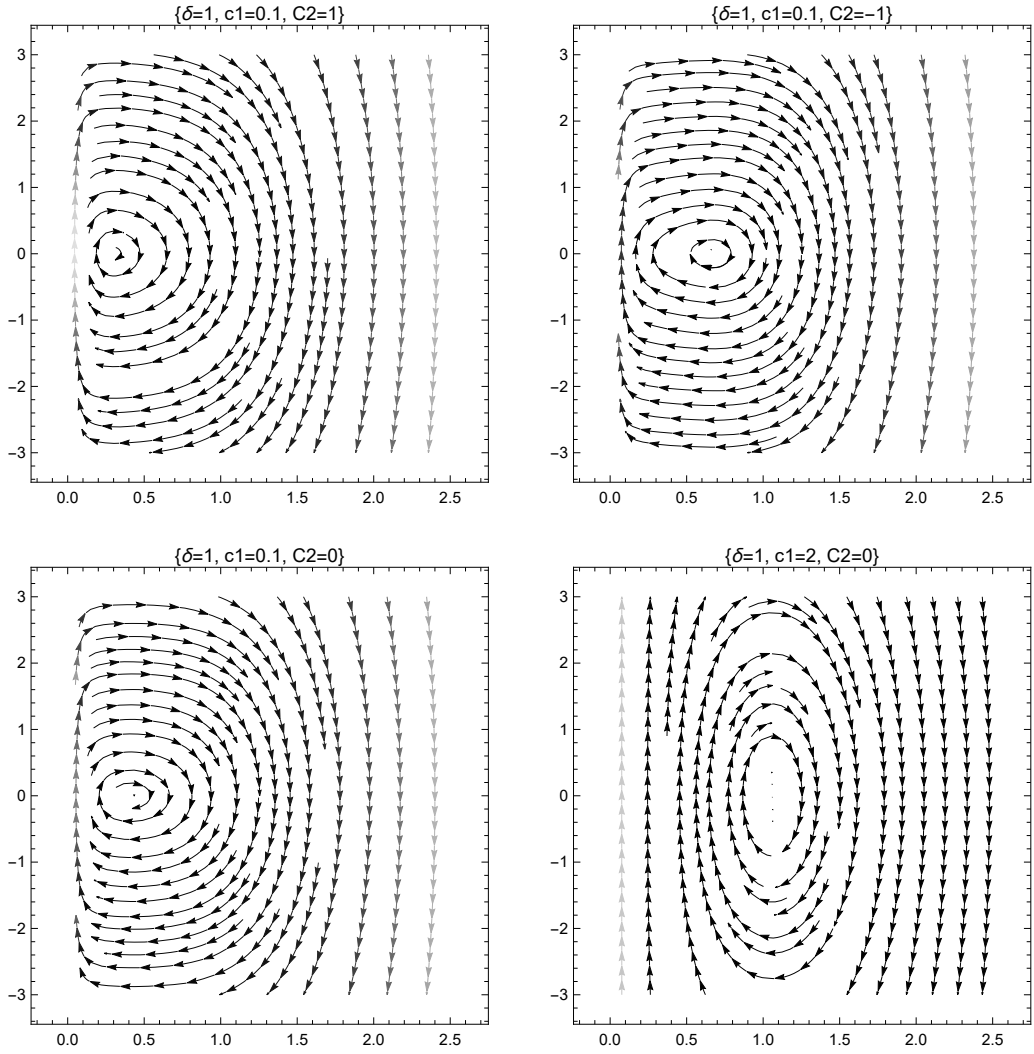}
\caption{Phase portrait for equation (\protect\ref{gc.14}) for various
values of the free parameters $\left\{ \protect\delta ,c_{1},\bar{C}%
_{2}\right\} $ and $\left( 1+\protect\alpha \protect\delta \right) =-1$. }
\label{fig2}
\end{figure}

In Fig. \ref{fig2} we present the phase portrait for the equation (\ref%
{gc.14}) and $\left( 1+\alpha \delta \right) <0,$ specifically for $\left(
1+\alpha \delta \right) =-1$, and for the rest of the constants to have the
same values as in Fig. \ref{fig1}. Again we observe a similar behaviour and
the existence of periodic solutions which correspond to travel-wave
solutions. 

As in the case of the static solution, the application of the singularity
analysis provides the analytic solution expressed by the Laurent expansion%
\begin{equation}
\varrho \left( x\right) =\varrho _{0}x^{-\frac{1}{2}}+\varrho _{1}x^{\frac{1%
}{2}}+\varrho _{2}x^{\frac{3}{2}}+...~,
\end{equation}%
where now%
\begin{equation}
\left( \varrho _{0}\right) ^{2}=\pm \frac{i}{\delta }~,~\varrho _{1}=-\frac{%
\left( -1\right) ^{\frac{3}{4}}}{2\delta ^{\frac{3}{2}}}\left( 1+\alpha
\delta \right) ~,~\varrho _{2}=-\frac{\left( -1\right) ^{1/4}\left( 9\left(
1+\alpha \delta \right) ^{2}-8\bar{C}_{2}\delta \right) }{24\delta ^{\frac{5%
}{2}}},~...~.
\end{equation}%
and $\varrho _{3}$ is an arbitrary constant.

\begin{figure}[tbp]
\centering\includegraphics[width=0.8\textwidth]{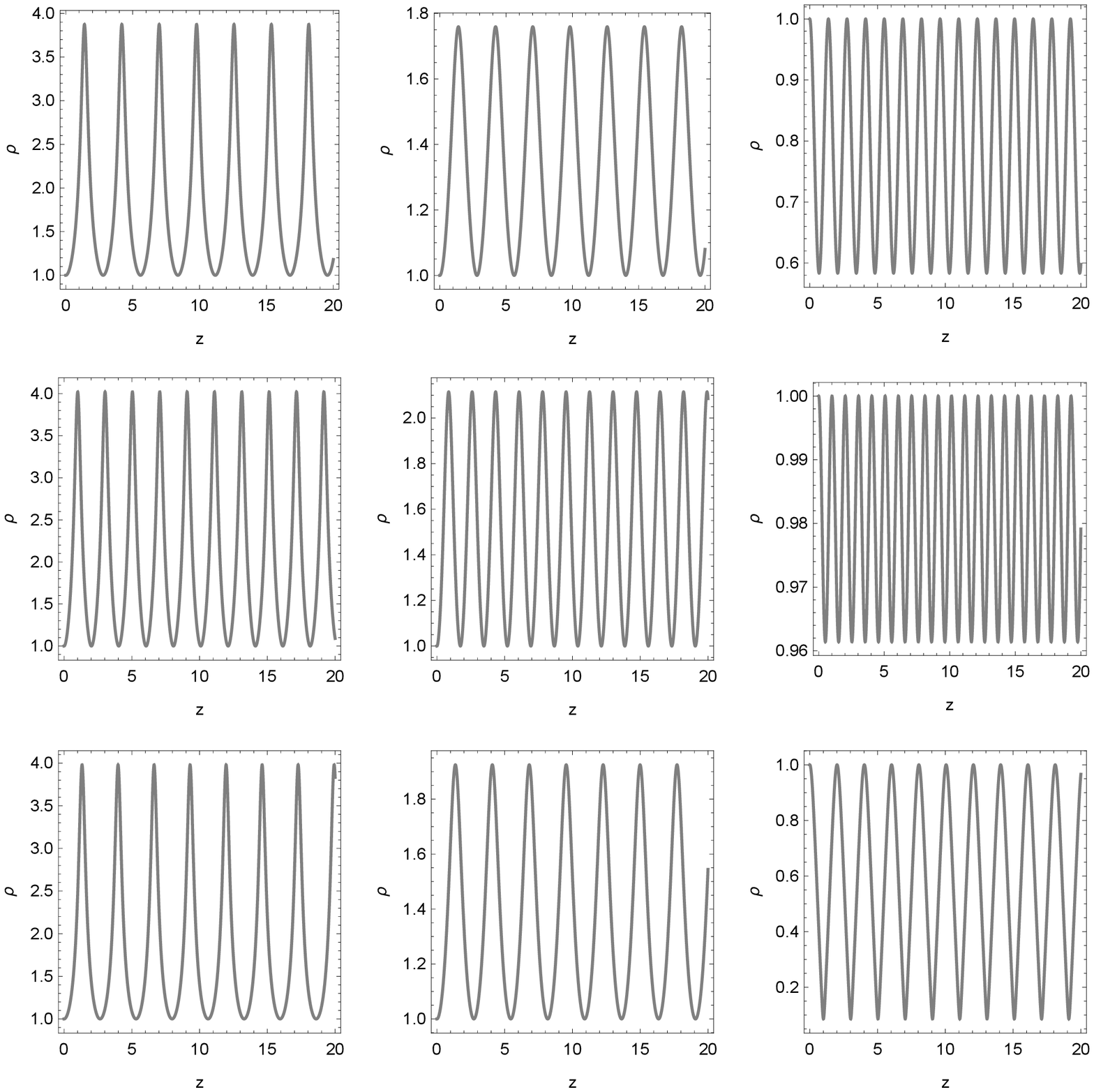}
\caption{Numerical simulation of equation (\protect\ref{gc1}) for various
values of the free parameters $\left\{ c_{1},C_{2},\left( 1+\protect\alpha 
\protect\delta \right) \right\} $ and initial conditions $\protect\varrho %
\left( 0\right) =1$ and~$\protect\varrho _{z}\left( 0\right) =0.$ The plots
of the first row are for (left to right) $\left\{ 1,1,0.5\right\} ,~\left\{
1,1,1\right\} ,~\left\{ 1,1,3\right\} $, the plots of the second row are for 
$\left\{ 2,0,0.5\right\} ,~\left\{ 2,0,1\right\} $,~$\left\{ 2,0,3\right\} $
and the plots of the third row are for $\left\{ 0.1,0.1,0.5\right\}
,~\left\{ 0.1,0.1,1\right\} $ and $\left\{ 0.1,0.1,3\right\} $. }
\label{nm1}
\end{figure}

In Fig. \ref{nm1} we present the numerical solution of (\ref{gc.14}) where
we observe the existence of travel-waves, i.e. a kink solutions, for the
gCLL equation.

\subsubsection{Lie symmetry $X_{3}$}

We proceed with the application of the similarity transformation provided by
the scaling symmetry $X_{3}$. \ The similarity transformation is 
\begin{equation}
u\left( t,x\right) =-\frac{1}{\delta }+\frac{1}{\sqrt{t}}U\left( \sigma
\right) ~,~\rho \left( t,x\right) =\frac{1}{\sqrt{t}}R^{2}\left( \sigma
\right) ~\text{with}~\sigma =\left( \frac{x}{\sqrt{t}}+\frac{\sqrt{t}}{%
\delta }\right) \text{.}
\end{equation}

Therefore, the reduced system after the application of the latter similarity
transformation is derived to be%
\begin{equation}
2RU-\sigma R+\delta R^{2}=c_{1},
\end{equation}%
\begin{equation}
R_{\sigma \sigma }-U^{2}R+\left( \sigma -2\delta R^{2}\right) UR+c_{2}R=0,
\end{equation}%
where $c_{1},~c_{2}$ are two integration constants.

Therefore, we end with only one differential equation%
\begin{equation}
R_{\sigma \sigma }+\frac{3}{4}\delta ^{2}R^{5}-2\left( 1+\alpha \delta
\right) R^{3}+\left( C_{2}+\frac{\sigma ^{2}}{4}\right) R-\frac{c_{1}^{2}}{%
R^{3}}=0  \label{gc1}
\end{equation}%
where now $C_{2}=\left( \frac{c_{2}}{2}-\delta c_{1}\right) $. The main
difference with the previous reduction is that the coefficient of the linear
term is not a constant.

\begin{figure}[tbp]
\centering\includegraphics[width=0.8\textwidth]{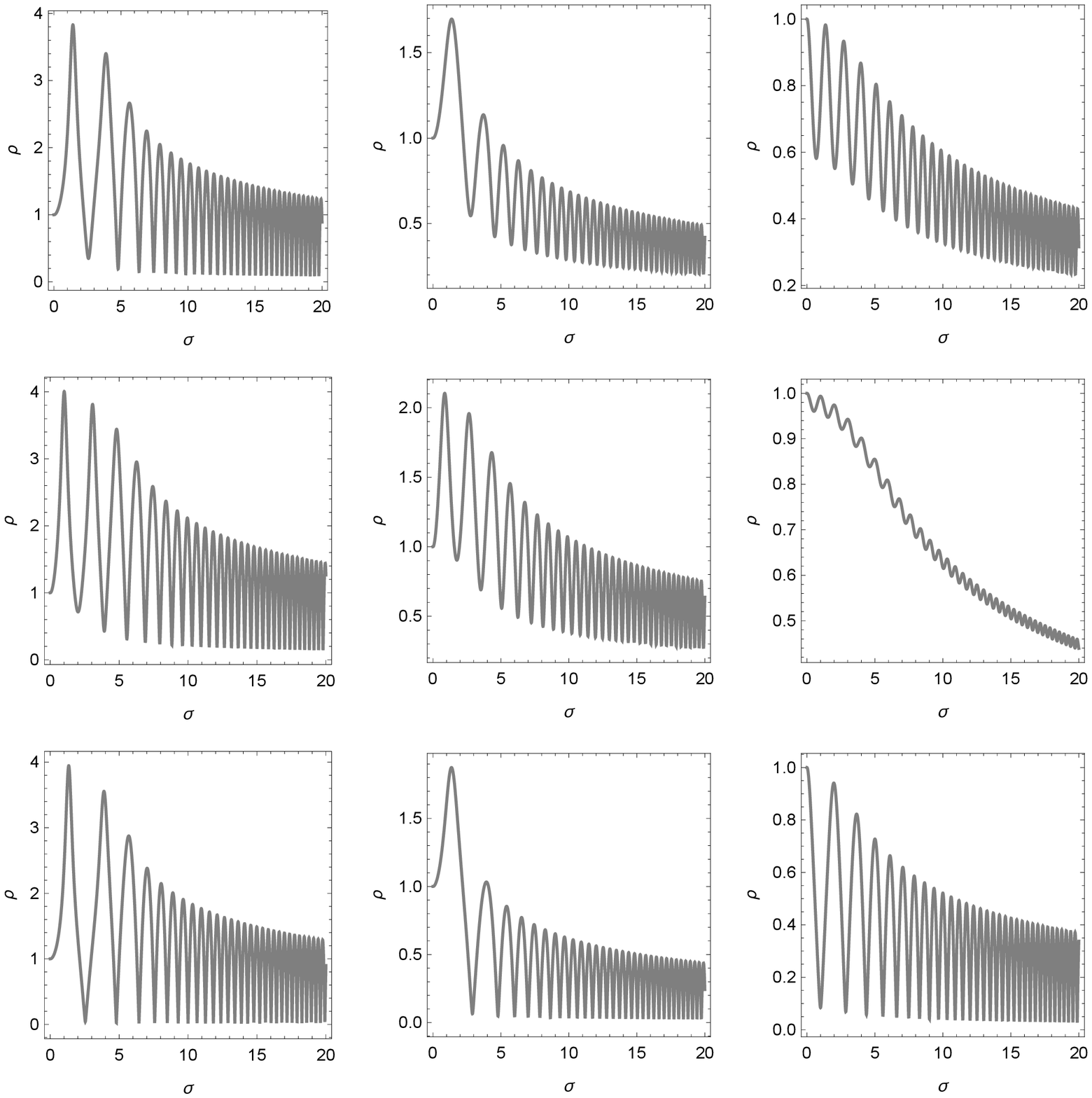}
\caption{Numerical simulation of equation (\protect\ref{gc1}) for various
values of the free parameters $\left\{ c_{1},C_{2},\left( 1+\protect\alpha 
\protect\delta \right) \right\} $ and initial conditions $R\left( 0\right)
=1 $ and~$R_{\protect\sigma }\left( 0\right) =0.$ The plots of the first row
are for (left to right) $\left\{ 1,1,0.5\right\} ,~\left\{ 1,1,1\right\}
,~\left\{ 1,1,3\right\} $, the plots of the second row are for $\left\{
2,0,0.5\right\} ,~\left\{ 2,0,1\right\} $,~$\left\{ 2,0,3\right\} $ and the
plots of the third row are for $\left\{ 0.1,0.1,0.5\right\} ,~\left\{
0.1,0.1,1\right\} $ and $\left\{ 0.1,0.1,3\right\} $.}
\label{fig3}
\end{figure}

Easily, we observe that the resulting solution is again periodic solution
but in this case around a central which moves. In Fig. \ref{fig3} we present
numerical simulation for equation (\ref{gc1}) for the initial condition $%
R\left( 0\right) =1~,~R_{\sigma }\left( 0\right) =0\,\ $and for various
values of the free parameters $\left\{ c_{1},C_{2},\left( 1+\alpha \delta
\right) \right\} $, from the figures it is clear the periodic behaviour of
the scaling solution, that is nothing else than a kink solution.

For small values of $R$, the dominant terms are the 
\begin{equation}
R_{\sigma \sigma }+\left( C_{2}+\frac{\sigma ^{2}}{4}\right) R-\frac{%
c_{1}^{2}}{R^{3}}=0
\end{equation}%
which is nothing else than the Ermakov-Pinney equation \cite{erm2,erm3,erm4}.

\section{Singularity analysis}

\label{sec4}

Let us now apply the ARS algorithm to study if the system (\ref{gc.02}), (%
\ref{gc.03}) possess the Painlev\'{e} property.

In order to determine the singular behaviour we replace in (\ref{gc.02}), (%
\ref{gc.03}) the following expression%
\begin{equation*}
\rho \left( t,x\right) =\rho _{0}\left( t,x\right) \phi \left( t,x\right)
^{p}~,~u\left( t,x\right) =u_{0}\left( t,x\right) \phi \left( t,x\right) ^{q}
\end{equation*}%
where $\phi \left( t,x\right) $ is a singular function. Hence, it follows
that the leading order terms provide $\left\{ p,q\right\} =\left(
-1,-1\right) $, where 
\begin{equation*}
\rho _{0}\left( t,x\right) =-\frac{1}{\delta ^{2}}\phi
_{x}^{2}~,~u_{0}\left( t,x\right) =-\frac{1}{4}\phi _{x}^{2}\text{.}
\end{equation*}

For the second step of the ARS algorithm, we determine the resonances which
are $\left\{ s_{1},s_{2},s_{3},s_{4}\right\} =\left\{ -1,2,2,3\right\} $
from where we conclude that the solution is given by Right Painlev\'{e}
Series. Indeed the solutions is expressed as 
\begin{equation}
\rho \left( t,x\right) =\rho _{0}\left( t,x\right) \phi \left( t,x\right)
^{-1}+\rho _{1}\left( t,x\right) +\rho _{2}\left( t,x\right) \phi \left(
t,x\right) +\rho _{3}\left( t,x\right) \phi \left( t,x\right) ^{3}+...~,
\end{equation}%
\begin{equation}
u\left( t,x\right) =u_{0}\left( t,x\right) \phi \left( t,x\right)
^{-1}+u_{1}\left( t,x\right) +u_{2}\left( t,x\right) \phi \left( t,x\right)
+u_{3}\left( t,x\right) \phi \left( t,x\right) ^{3}+...~.
\end{equation}%
where the consistency test gives that functions $\rho _{2}\left( t,x\right)
,~\rho _{3}\left( t,x\right) $ and $u_{2}\left( t,x\right) $ are arbitrary.

We conclude that the generalized Chen-Lee-Liu equation possess the Painlev%
\'{e} property and its solution is expressed by Right Laurent expansions.
The latter expressions can describe Kink solutions for specific initial
conditions.

\section{Conclusion}

\label{sec5}

In this work we determined exact and analytic periodic solutions for the
gCLL equation with the use of Lie symmetries. The gCLL equation admits a
three-dimensional Lie algebra, which leads to four different similarity
transformations. In particular we found static, stationary, travel-wave and
scaling similarity solutions.

Except from the stationary solution, which is not of interest, the other
solutions are periodic solutions. It is interesting that in all cases for
small values of the intensity variable $\rho \left( t,x\right) $ the gCLL is
reduced to the Ermakov-Pinney equation, which is a well-known integrable
system. \ 

Moreover, we investigated if the gCLL possess the Painlev\'{e} property, for
that we applied the ARS algorithm where we were able to write the analytic
solution of the gCLL with the use of Right Painlev\'{e} Series.

The periodic solutions, provided by the similarity transformations, are
direct related with the existence of dark and bright solitons for the
nonlinear differential equation \cite{so1,so2,so3}. Optical solitons are
exact solutions of mathematical models with direct applications in the
information transfer in optical fibers \cite{so4}. As far the results or our
analysis, are concerned, the travel-wave solution which was found before for
the gCLL describes a 1-soliton solution known as kink solution. Furthermore,
the scaling solution is also a kink solution, where now the amplitude of the
oscillation is not a constant. 

The determination of these new kink solutions is essential for the physical
viability of the model. A study of the properties of the kink solutions and
their real world applications extends the scopus of this study and will be
performed in a future study. 

\appendix

\section{Lie symmetries}

\label{ap1}

We briefly discuss the main definition and algorithm for the determination
of Lie point symmetries. Consider the one-dimensional parameter point
transformation \ 
\begin{align}
\bar{x}^{k}& =x^{k}+\epsilon \xi ^{i}\left( x^{k},u\right) , \\
\bar{\eta}& =\eta +\epsilon \eta \left( x^{k},u\right) .
\end{align}%
with generator $X=\xi ^{i}\left( x^{k},u\right) \partial _{i}+\eta \left(
x^{k},u\right) \partial _{u}$, then the differential equation $\mathbf{H}%
\left( x^{k},u,u_{i},u_{ij},...,u_{i_{1}i_{2}...i_{n}}\right) $ is invariant
under the action of the one parameter point transformation if and only if%
\begin{equation*}
\lim_{\varepsilon \rightarrow 0}\frac{\mathbf{\bar{H}}\left( \bar{x}^{k},%
\bar{u},\bar{u}_{i},\bar{u}_{ij},...,\bar{u}_{i_{1}i_{2}...i_{n}}\right) -%
\mathbf{H}\left( x^{k},u,u_{i},u_{ij},...,u_{i_{1}i_{2}...i_{n}}\right) }{%
\varepsilon }=0,
\end{equation*}%
or equivalently, if there exists a function $\lambda $ such that the
following condition to be true \cite{ibra,Bluman,Stephani}%
\begin{equation}
X^{\left[ n\right] }\mathbf{H}-\lambda \mathbf{H=0}
\end{equation}%
where $X^{\left[ n\right] }$ is called the n-th prolongation/extension of $%
X~ $in the jet-space defined as%
\begin{equation}
X^{\left[ n\right] }=X+\left( D_{i}\eta -u_{,k}D_{i}\xi ^{k}\right) \partial
_{u_{i}}+\left( D_{i}\eta _{j}^{\left[ i\right] }-u_{jk}D_{i}\xi ^{k}\right)
\partial _{u_{ij}}+...+\left( D_{i}\eta _{i_{1}i_{2}...i_{n-1}}^{\left[ i%
\right] }-u_{i_{1}i_{2}...k}D_{i_{n}}\xi ^{k}\right) \partial
_{u_{i_{1}i_{2}...i_{n}}}.
\end{equation}

If $X$ is a symmetry vector for the differential equation $\mathbf{H}$, then
we can always find a coordinate transformation such that the symmetry vector
to be written in the canonical coordinates, i.e. $X=\partial _{x^{n}}$,
where the differential equation is%
\begin{equation*}
\mathbf{H=H}\left( x^{\mu },u,u_{i},u_{ij},...,u_{i_{1}i_{2}...i_{n}}\right)
~,~\mu \neq n,
\end{equation*}%
clear from the last expression it follows $\partial _{x^{\mu }}H=0$. The
coordinate transformation which leads to the canonical coordinates it is
called similarity transformation and it is mainly applied for the reduction
of the differential equation,

\section{One-dimensional optimal system}

\label{ap2}

Let a given differential equation admit as Lie symmetries the elements $%
\left\{ X_{1},~X_{2},~...~X_{n}\right\} $ of the $n$-dimensional Lie algebra 
$G_{n}$ with structure constants $C_{jk}^{i}$. The two symmetry vectors $%
Z,~W $ defined as 
\begin{equation}
Z=\sum_{i=1}^{n}a_{i}X_{i}~,~W=\sum_{i=1}^{n}b_{i}X_{i}~,~\text{\ }%
a_{i},~b_{i}\text{ are constants.}  \label{sw.04}
\end{equation}%
we shall say that are equivalent if and only if \cite{olver} 
\begin{equation}
\mathbf{W}=\sum_{j=i}^{n}Ad\left( \exp \left( \epsilon _{i}X_{i}\right)
\right) \mathbf{Z}  \label{sw.05}
\end{equation}%
or 
\begin{equation}
W=cZ~,~c=const\text{ that is }b_{i}=ca_{i}\text{.}  \label{sw.06}
\end{equation}

Operator $Ad\left( \exp \left( \epsilon X_{i}\right) \right) X_{j}$ defined
as 
\begin{equation}
Ad\left( \exp \left( \epsilon X_{i}\right) \right) X_{j}=X_{j}-\epsilon 
\left[ X_{i},X_{j}\right] +\frac{1}{2}\epsilon ^{2}\left[ X_{i},\left[
X_{i},X_{j}\right] \right] +...,  \label{sw.07}
\end{equation}%
is called the adjoint representation.$~$

The determination of all the one-dimensional subalgebras of $G_{n}$ which
are not related through the adjoint representation it is necessary in order
to perform a complete classification of all the possible similarity
transformations, i.e. similarity solutions, for a given differential
equation. This classification is known as the one-dimensional optimal system.

\section{Singularity analysis}

\label{ap3}

The development of the Painlev\'{e} Test for the determination of
integrability of a given equation or system of equations and its
systematization been succinctly summarized by Ablowitz, Ramani and Segur in
the so-called ARS algorithm \cite{Abl1,Abl2,Abl3}. \ The ARS algorithm is
constructed by three basic steps, they are: (a) determine the leading-order
term which describes the behaviour of the solution near the singularity, (b)
find the position of the resonances \ which shows the existence and the
position of the integration constants and (c) write a Laurent expansion with
leading-order term determined in the first step in order to perform the
consistency test and the solution, for a review on the ARS algorithm and
various applications we refer the reader in \cite{sinn1}, while in \cite%
{sinn2} a discussion between the Lie's approach and the singularity analysis
is given.

\end{document}